\documentclass[proceedings, preprint]{rmaa}

\usepackage{paralist}

\usepackage{psfrag,color}

\def\msun{\mbox{M$_{\odot}$}}  

\def\liso{\mbox{L$_{\rm iso}$}}
\def\eiso{\mbox{E$_{\rm iso}$}}
\def\ngrb{\mbox{$\dot{n}_{\rm GRB}(z)$}}
\def\rsfr{\mbox{$\dot{\rho}_{\rm SFR}(z)$}}

\def\fldl{\mbox{$f$(L$_{\rm iso}$)dL$_{\rm iso}$}}

\def\ergs{\mbox{erg \ s$^{-1}$}}
\def\nz{\mbox{$\dot{N}(z)$}}
\def\np{\mbox{$N(P)$}}
\def\ep{\mbox{$E_{\rm pk}$}}
\def\tdur{\mbox{t$_{\rm 45}$}}

\def\Om{\mbox{$\Omega_{\rm m}$}}
\def\OL{\mbox{$\Omega_{\Lambda}$}}
\def\wo{\mbox{$w_0$}}

\def\sw{\mbox{\it Swift}}

\def\mathnew{\mathsurround=0pt}
\def\ref{\par\noindent\hangindent=2pc \hangafter=1 }

\def\simov#1#2{\lower .5pt\vbox{\baselineskip0pt

\lineskip-.5pt\ialign{$\mathnew#1\hfil##\hfil$\crcr#2\crcr\sim\crcr}}}

\def\simless{\mathrel{\mathpalette\simov <}}

\SetYear{2007}
\SetConfTitle{XII IAU Regional Latinamerican Meeting}

\title{Gamma-Ray Bursts, new cosmological beacons} 

\author{
  V. Avila-Reese,\altaffilmark{1} 
  C. Firmani,\altaffilmark{1,2}
  G. Ghisellini\altaffilmark{2}
  and J. I. Cabrera\altaffilmark{1}}

\altaffiltext{1}{Instituto de Astronom\'ia, Universidad Nacional Aut\'onoma de M\'exico,       A.P. 70--264, 04510, M\'exico, D.F.}

\altaffiltext{2}{INAF--Osservatorio Astronomico di Brera, via E.Bianchi 46, I--23807 Merate, Italy}

\shortauthor{Avila-Reese et al.}
\shorttitle{GRBs, new cosmological beacons}

\listofauthors{V. Avila-Reese et al.}
\indexauthor{Avila-Reese, V.}
\indexauthor{Firmani, C.}
\indexauthor{Ghisellini, G.}
\indexauthor{Cabrera, J. I.}

\abstract{Long Gamma-Ray Bursts (GRBs) are the brightest electromagnetic explosions
in the Universe, associated to the death of massive stars.  As such, GRBs
are potential tracers of the evolution of the cosmic massive star formation, 
metallicity, and Initial Mass Function. GRBs also proved to be appealing cosmological 
distance indicators. This opens a unique opportunity to constrain the cosmic 
expansion history up to redshifts 5--6. A brief review on both subjects is presented here. }

\resumen{Los Estallidos de Rayos Gamma (ERG's) largos son las explosiones electromagn\'eticas
m\'as potentes del Universo, asociadas a la muerte de estrellas masivas. Como tales, los ERG's
son trazadores potenciales de la evoluci\'on de la formaci\'on estelar c\'osmica,
la metalicidad y la funci\'on inicial de masa. Los ERG's tambi\'en han probado ser atractivos
como indicadores de distancia cosmol\'ogicos, lo cual abre una oportunidad \'unica de
constre\~nir la historia de expansi\'on c\'osmica hasta $z\approx 5-6$.
Se presenta aqu\'{\i} una rese\~na sobre ambos temas. }

\addkeyword{Cosmology: observations}
\addkeyword{gamma--rays: bursts}
\addkeyword{stars: star formation history}

\begin{document}
\maketitle

\section{Introduction}
\label{sec:intro}

Detected as brief ($\sim 0.01-100$ s), intense flashes of $\gamma$--rays (mostly sub--MeV),
Gamma--ray Busts (GRBs) are the brightest electromagnetic explosions in the Universe. 
The power emitted by GRBs in 
electromagnetic form can reach luminosities up to $L\sim 10^{52}-10^{53}$ \ergs, while
AGNs can have $L\sim 10^{48}$ \ergs~ (but for long times), 
and Supernovae can have $L\sim 10^{45}$ \ergs~ for the first hundreds of seconds after the 
explosion. The short variability timescales of the $\gamma-$ray emission, suggest already
very small dimensions for the sources, of the order of tens of kilometers, typical of stellar
black holes or neutron stars. Several pieces of evidence indeed show that GRBs are associated 
with cataclysmic stellar events, and that the $\gamma-$ray emission comes from highly 
relativistic collimated outflows. The typical bulk Lorentz factor for the jets
is $\Gamma\sim 300$. Thus, GRBs are true cosmic laboratories for the study
of relativistic, magneto--hydrodynamical, and high energy processes (for recent reviews on
the GRB physics see e.g., Zhang \& M\'esz\'aros 2004; Piran 2005; M\'esz\'aros 2006). 

Furthermore, GRBs and their afterglows are of great interest for studies related to stellar 
astrophysics, the interstellar and intergalactic medium, and most important, they reveal 
themselves as unique probes of the high redshift Universe. In the last 3 years, on average 
$\sim 1$ paper per day is published on GRBs in refereed journals and $\sim 4$ per day in 
non--refereed publications.
 
GRBs are divided into two main groups which, following the notation of Zhang (2007), 
we will call Type I and Type II. The former ($\approx 1/3$) have short $\gamma-$ray durations
($\lesssim 2$ s) and hard spectrum; it is conjectured that they result from binary mergers
of compact stellar objects (NS--NS or NS--BH). The latter ($\approx 2/3$) have durations
larger than 2 s and their $\gamma-$ray spectra tend to be softer. The observations show 
(e.g., Hjorth et al. 2003; Stanek et al. 2003) that these GRBs result from the collapse of 
rapidly rotating cores of low--metallicity stars more massive than about 25 \msun~ 
('collapsar' scenario). 

A breakthrough in the GRB field happened a decade ago: thanks to their relatively long duration, 
for some Type II GRBs it was possible to resolve 
and detect the afterglow at softer (X--ray, optical, IR and radio) energies. This allowed to 
measure spectral lines and/or to identify the host galaxy; hence the redshifts, $z$, could be 
determined. Up to December 2007, there were around 100 GRBs with secure $z$ measurements 
(in $\sim 90\%$ of the cases by the afterglow and $\sim 10\%$ of the cases by the host galaxy). 
More than 60\% of the  $z$ determinations were obtained during the last 3 years with the 
dedicated \sw~ satellite (Gehrels et al. 2004), which allowed also to discover the afterglows 
for some Type 
I events (7 up to date). No doubt, GRBs are the brightest transient cosmological events
measurable. In Fig. 1 the $z$ distribution of \sw~ GRBs is shown: it ranges from $z=0.033$ to 6.29, 
with an average $z$ of 2.2. Indirect estimates suggest, however, that many of the observed Type 
II GRBs could be produced at $z'$s larger than 6 (e.g., Bromm \& Loeb 2005).

Summarizing: type II GRBs are extremely powerful explosions associated to the collapse of 
short--lived massive stars, with peak emission at sub--MeV energies, where dust
extinction is not an issue. Besides, the $\gamma-$ray spectra of type II GRBs
are such that the $K$ correction is small or even negative. Thus, the fluxes of these 
events can be detected eventually from any $z$. All these properties convert type II GRBs in 
cosmological beacons, which can help us look back in time. We will review
here advances along this line in two directions: GRBs\footnote{Hereafter GRB refers to
a type II burst.} as tracers of the history of global 
massive star formation rate (SFR; \S 2), and GRBs as cosmic rulers able to help us
in constraining the expansion history of the Universe (\S 3). In \S 4, perspectives
for future work will be discussed.

\section{GRBs as tracers of the global massive star
formation rate}
\label{sec:SFH}

The death rate of massive short--lived stars resembles their formation rate. 
Thus, the GRB formation rate (GFR) can be used as a potential tracer of the massive 
SFR in the Universe. Current observations allow to construct the (yet very incomplete)
GRB $z$ distribution per unit of time, \nz. This {\it observable} distribution 
is connected to the history of the {\it intrinsic} GFR (per unit of comoving volume), 
\ngrb, through:
\begin{equation}
\nz=\int_0^{z}dz\frac{dV/dz}{1+z}\frac{F(z)}{f_{obs}}\ngrb,
\end{equation}
where $dV/dz$ is the comoving volume element, $(1+z)^{-1}$ accounts for the time dilation,
$f_{obs}$ stands for the detector exposure factor and the average GRB beaming,
and $F(z)=\phi_{flux}(z)\phi_z(z)$ takes into account several selection effects. 
$F(z)$ can be understood as the probability to detect the burst and its afterglow, 
and to measure its $z$ from the afterglow; it comprises two classes of effects: 
\begin{asparaitem}
\item the flux--limited selection function, 
\begin{equation}
\phi_{flux}(z)=\int_{\liso(P_{\rm min},z)}^{\infty}\fldl, 
\end{equation}
which depends on the detector 
flux threshold $P_{\rm min}$ and on the GRB luminosity function (LF), \fldl~ (\liso~ is the 
equivalent isotropic --beam uncorrected-- GRB bolometric luminosity).

\item the selection function related to the observability of the afterglow and its $z$ 
determination, $\phi_z(z)$. This function is very uncertain and
comprises spectroscopic and photometric selection effects (e.g., observability of lines 
in the spectral range and in the presence of night--sky lines, afterglow flux--limited
selection, different systematics in the emission or absorption line technique 
used to determine $z$), as well as potential astrophysical effects (e.g., obscuration, 
evolving dust extinction, a bias in the GRB host galaxy population). An evidence that 
these selection effects are severe is the fact that only $\sim 1/3$ of the
 \sw~ GRBs have reliable $z$ determinations in spite of the quick afterglow localization
and the great effort in follow--up observations.
\end{asparaitem}

\begin{figure}[!t]
  \includegraphics[angle=-90,width=\columnwidth]{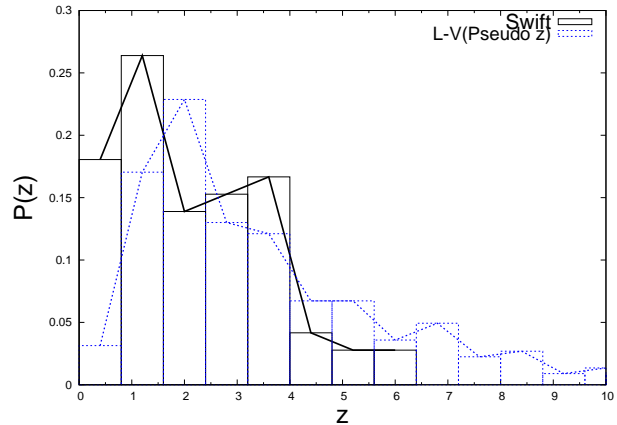}
  \caption{Normalized distribution of 72 spectroscopically measured
$z'$s obtained with \sw~ (solid line) and 220 pseudo--redshifts
obtained with the \liso--variability correlation (dashed line).}
  \label{fig:simple}
\end{figure}

\subsection{Direct inferences, yet limited}

 A direct joint determination of \ngrb~ and \fldl~ would be possible from a populated 
$\liso-z$ diagram. However, the situation is complicated mainly by some of the uncertain
component biases present in the $\phi_z(z)$ function (e.g., Bloom 2003; Gou et al. 2004; 
Fiore et al. 2007; Coward et al. 2007). In view of this and the
yet low population statistics, it is not feasible to attempt a clean reconstruction 
of \ngrb~ from current data. Recent works that make use of the \sw~ data start from some 
assumptions and limit themselves to explore only statistical compatibilities of the observed 
\nz~ distribution with models of \ngrb~ assumed proportional to the global SFR history
inferred from extragalactic studies, \rsfr. A general conclusion is that \nz~ implies 
an enhanced \ngrb~ at high redshifts with respect to 
 \rsfr~ (Kistler et al. 2008; see also Le \& Dermer 2007; Guetta \& Piran 2007).
It should be noted that some pieces of evidence suggest that the \sw~ sample of GRBs with 
$z$ determined is a fair sample of the real high$-z$ GRB population (Fiore et al. 2007).
 
On the other hand, Coward et al. (2007) concluded that the observed \sw~ $z-$distribution,
where the number increasing from $z\approx 0$ to $z\approx 1.5$ is modest, 
implies a bias in the afterglow observability such that at $z=0-1$ it works inversely 
proportional to the global SFR history. It could be that the \sw~ sample with $z$ is biased 
against low--$z$ GRBs, probably due to the enhanced extinction associated with the prolific SFR at
$z\sim 1$ (Fiore et al. 2007; Coward et al. 2007). 

\subsection{Indirect inferences, encouraging results}

Alternatively to the observed $\liso-z$ diagram, there are other methods for
inferring the GFR history from observations but in a model--dependent way. 
We will mention three methods:

(1) The most extensive GRB observational database is the CGRO-$BATSE$ peak flux $P$ 
distribution, \np, for $\sim 3000$ bursts. The \np~ distribution (corrected by
the exposure factor) is the result of 4 physical ingredients: \ngrb, \fldl, the jet 
opening angle distribution, $f(\theta_j$), and the volumetric factor given by the 
cosmological model. Therefore, the inference 
of \ngrb~ from the observed \np~ is a highly degenerated problem. The adequate 
introduction of complementary observational information helps to overcome partially 
the degeneracies; for example, the \ep~ distribution (Daigne et al. 2006) or the
$\theta_j$ distribution (Le \& Dermer 2007). 
 
(2) The $\liso-z$ diagram can be (indirectly) obtained for large data-sets of GRBs
without $z$ measured by using empirical correlations that involve \liso. For example,
Lloyd--Ronning et al. (2002) and Yonetoku et al. (2004) inferred \ngrb~ by applying the 
\liso--variability correlation (Fenimore \& Ramirez--Ruiz 2000) to 220 BATSE GRBs,
and the $\liso-E_{\rm pk,rest}$ relation to 689 BATSE GRBs, respectively. This method
relies on the certainty and accuracy of the used empirical correlation.

(3) Given an adequate parametrization for \ngrb~ and \fldl, their free parameters
can be efficiently constrained by fitting models 
{\it jointly} to both the observed \np~ distribution and the \nz~ distribution
inferred as in (2) or as in \S\S 2.1 after selection effects correction. So far, 
this method is the most powerful. A description of and the results from this method  
(Firmani et al. 2004; hereafter FAGT) are as follows.

{\bf The method.} The observed \np~ and \nz~ distributions are modeled by seeding at 
each $z$ a large number of GRBs with a given rate, \ngrb, and LF, \fldl, and then by 
propagating the flux of each source to $z=0$. \liso\ in 
the rest frame is defined as \liso=$\int_{\rm 30keV}^{\rm 10000keV}ES(E)dE$,
where $S(E)$ is the Band (Band et al. 1993) energy spectrum. The break energy at rest, 
$E_b$, is assumed either constant (512 keV) or dependent on 
\liso~ according to the ``Yonetoku relation'' (Yonetoku et al. 2004).
FAGT explored two models for \fldl, the single and double power laws (SPL
and DPL, respectively), and two cases, one where \fldl~ is constant in 
time, and another one where \liso\ evolves as $(1+z)^{\delta}$. The
function \ngrb~ was modeled as 
\begin{equation}
\ngrb=K\times e_{\rm SFR}(z;a,b)\times \eta_{z>2}(z;c),
\end{equation}
where $e_{\rm SFR}(z;a,b)$ is a bi--parametric function proposed by Porciani \& Madau 
(2000) for fitting the observed SFR history, and $\eta_{z>2}(z;c)$ allows to control the 
growth or decline of \ngrb~ at $z>2$.  The idea is to constrain the parameters of
the LF (2 and 3 for the SPL and DPL, and 1 more if evolution is allowed ) 
and of \ngrb~ (4 parameters) by applying a {\it joint} fit of the model predictions to the 
\np~ and \nz~ distributions. 
The \np~ distribution for more than 3000 GRBs compiled by
Stern et al. (2002) and the $z$ distribution inferred for 220 GRBs in Lloyd--Ronning
et al. (2002; see above) were used. 

{\bf Results.} FAGT obtained that an evolving LF is preferred in all the analyzed
cases (with an optimal value of $\delta=1\pm 0.2$). For non evolving LFs, the predicted 
\np~ distributions have systematically an excess of GRBs at the bright end, and the peak 
of the \nz~ distribution is shifted to higher $z'$s (see their Fig. 1). Only a weak 
preference was found for the SPL model over the DPL one. On the other hand, the best
fits not only imply an evolving LF but also GFRs histories that steeply increase (by a 
factor of $\sim 30$) from $z=0$ to $z\approx 2$ and then continue increasing gently up 
to $z\sim 10$ as $(1+z)^{1.4}$ and $(1+z)$ for the SPL and DPL LFs, respectively. 

Figure 2 reproduces from FAGT the SFR history as traced by the GFR history (dot--shaded region)
under an opportune normalization and assuming a constant initial mass function (IMF). 
Both the best SPL and DPL LF models with evolution ($\delta = 1$) and their $1\sigma$ 
uncertainties are 
taken into account by the region. Symbols are the SFR traced by the rest--frame UV luminosity 
and corrected for dust obscuration (Giavalisco et al. 2004), while the solid line is
a more recent piecewise fit to a large compilation and dust correction of SFR observations by 
Hopkins \& Beacom (2006). The main difference between the SFR histories inferred from GRBs 
and from extragalactic observations is the enhancement (that increases with $z$) of the former
 with respect to the latter after $z= 1-2$. Interesting enough, the \sw~ based
 studies mentioned in \S\S 2.1, while in much less detail, attain a similar conclusion.

\begin{figure}[!t]
  \includegraphics[width=\columnwidth]{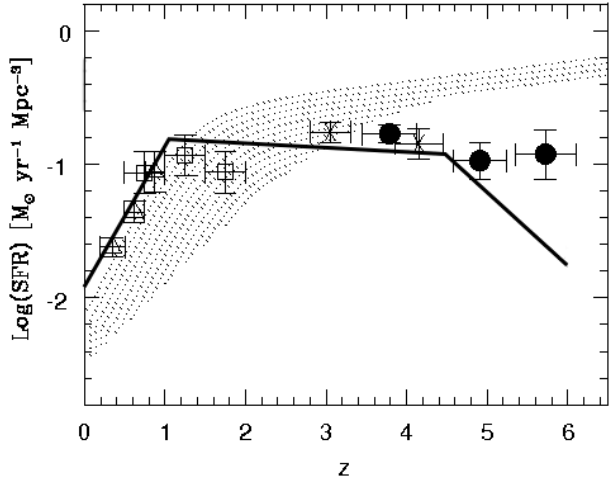}
  \caption{The cosmic SFR histories as traced by GRBs (dotted region) 
and by rest--UV luminosity (symbols and solid line, the latter from  the 
recent fit to several observations by Hopkins \& Beacom 2006); see text. }
  \label{fig:simple}
\end{figure}

{\bf Implications.} 
Why the GFR evolution might be enhanced with respect to the rest--UV traced SFR evolution 
at high redshifts? Let us discuss the pros and cons of some of the possibilities.

\begin{asparaitem}
\item Since the $\gamma-$rays do not suffer absorption, even in very dense molecular regions,
GRBs are expected to trace the massive SFR of any galaxy/region in the Universe, for
example of the dust--enshrouded actively star--forming galaxies at high redshifts.
However, there are pieces of evidence against GRBs being located in this kind
of galaxies (e.g., Le Floc'h et al. 2007).
  
\item  Most of the GRB host galaxies are faint, low--mass, low--metallicity star-forming galaxies
(e.g., Fynbo et al. 2003; Stanek et al. 2006) and the GRB--to--SN ratio might increase significantly
at lower metallicities (Yoon et al. 2006). Therefore, the GFR could be tracing the SFR of
a biased population of low--metallicity galaxies. Nuza et al. (2007), by means of
cosmological simulations, have found indeed that the GRB host properties are reproduced if GRBs form
from low--metallicty progenitor stars. However, recent observational studies are showing that 
the neutral ISM around GRBs is not metal poor and is enriched by dust (see Savaglio 2007). 
Furthermore, GRB hosts should not to be special, but normal, faint, star-forming galaxies (the most
abundant), detected at any $z$ just because a GRB event has occurred. 

\item  If GRBs are produced
in binary systems and the probability of interloper--catalyzed binary mergers in dense
star clusters (where most GRBs appear to occur, Fruchter et al. 2006) increases with $z$,
then an enhancement in the GFR is expected (Kistler et al. 2007). This scenario requires 
quantitative calculations. Interesting enough, FAGT, and more recently Bogomazov, Lipunov
\& Tutukov (2007), have calculated the Galaxy production rates of Wolf--Rayet stars in 
very close binaries (progenitors of rapidly rotating --Kerr-- black holes), and found 
an approximate agreement with the estimated rates of type II GRBs in galaxies.

\item  An IMF that becomes increasingly top--heavy at higher $z$, would increase
the relative number of massive stars produced. Since the rest--UV luminosity
already traces massive SFR, the IMF should be biased to very massive stars 
(GRB progenitors) so that an evolving IMF could explain the difference
seen in Fig. 2.   
\end{asparaitem}

 The above discussion is an example of the large potential of GRB studies for
exploring such key issues in astronomy as SF at high $z$, and the
IMF and chemical evolution in galaxies. 

Regarding low $z'$s, Fig. 2 shows that the GFR decays with time closely proportional 
to the SFR since $z\approx 1$. This feature does not seem to be observed in the \sw~ $z$ 
distribution, which would imply
a dominant redshfit bias, $\phi_z(z)$ as suggested by Coward et al. (2007). In the
redshift range from $z=1$ to $z=2$, the SFR attains a broad maximum, while the 
GFR keeps increasing rapidly. The \sw~ $z$ distribution would suggest the opposite.
In fact, as shown in Fig. 1, the \sw~ and the inferred from the $L-$variability
relation \nz~ distributions differ quite substantially up to $z\approx 2$ (rather than
the absolute values, the important comparison between the two histograms is at the level 
of the fractional changes with $z$).  As remarked by Fiore et al. (2007), the present
samples with $z$ determinations seem to be largely incomplete, especially at $z\simless 2$.

\section{GRBs as tracers of the Universe expansion rate history}
\label{sec:EPS}

The energetics of GRBs spans 3--4 orders of magnitude; at first sight GRBs are all but 
standard candles. A breakthrough in the field happened after the discovery of a tight
correlation between the collimation corrected bolometric energy, 
$E_{\gamma}=E_{\rm iso}(1-\cos\theta_j)$
and the prompt peak energy in the $\nu f_{\nu}$ spectrum, \ep~ (Ghirlanda et al. 2004a). 
This and any other similar correlations allow to standardize the GRB energetics for using
GRBs as distance indicators in the Hubble diagram (HD). Such an endeavor, however, is 
not trivial. 

The first conceptual problem is that most of the GRBs with $z$ measured are at 
cosmological distances (see Fig. 1). Therefore, the given correlation can not be 
calibrated locally; to establish the correlation, a cosmology should be assumed, but 
the cosmological parameters are just what we pretend to constrain in the HD. By using
statistical approaches, this {\it circularity} problem can be treated in order to get optimal
constraints for the explored cosmological parameters. The idea is to use the best--fitted 
correlation (smallest scatter) for each cosmology and to find which cosmology produces
the smallest $\chi^2$ in the HD, constructed by applying the corresponding correlation.
A powerful Bayessian--like method to carry out this undertaking has been introduced by 
Firmani et al. (2005,2007). Other groups have developed alternative variants 
(see e.g., Xu, Dai  \& Liang 2005; Schaefer 2007; Li et al. 2007). On the other hand, 
Ghirlanda et al. (2006) have shown that in order to calibrate for example the ``Ghirlanda'' 
correlation it is enough to have a dozen of GRBs in a narrow ($\Delta z/z\sim 0.1$) redshift 
bin, something that will be possible in the near future.

{\bf Results.} The first cosmological results obtained by using the ``Ghirlanda'' correlation were
encouraging: they provided a test of the accelerated expansion independent from the SN Ia 
studies (Ghirlanda et al. 2004b; Firmani et al. 2005; see also Dai, Liang \& Xu 2004). The addition of
GRBs to the SN probe, reduces the confidence levels of the constrained cosmological parameters.
A drawback of the ``Ghirlanda'' correlation is that it needs to establish expensive follow--up 
observations: $\theta_j$ is determined from the achromatic break time, $t_{\rm break}$, in the 
afterglow light curve. 

Firmani et al. (2006a) discovered a tight correlation among three prompt 
emission GRB parameters: \liso, \ep, and duration \tdur. Some of the cosmological
constraints obtained by using the ``Firmani'' correlation are plotted in Fig. 3. 
We show in these plots how the confidence levels given by the SNLS survey (Astier et al. 2006) 
are improved when the GRB constraints are added. Right panel of Fig. 3 shows the HD for 117 SNLS 
supernovae and 19 GRBs by using the 'vanilla' $\Lambda$CDM cosmology (solid line), which 
provides a good 
fit to the observations. GRBs are a natural extension of SNIa to high $z'$.  In the bottom panel 
the residuals to the assumed cosmology are plotted; the averages and its uncertainties are
$0.15\pm 0.01$ mag and $0.26\pm 0.05$ mag for the SNe and GRBs, respectively. 

Our results in general (constraining only two parameters at the same time) showed that
the flat $\Lambda$CDM cosmology is consistent at the $1\sigma$ level with the HD of GRBs 
(and GRBs+SNe) up to $z=4.5$.  

The cosmography with GRBs opens a valuable window for exploring the expansion history of the
Universe to $z>2$ (more than $\sim$10 Gyr ago), where SNIa are practically impossible to 
observe. Besides, GRBs offer
some important advantages for cosmographic studies. (1) GRBs are not affected by dust 
extinction. (2) The luminous distance $d_L$ in the HD is a cumulative quantity with $z$, 
so that the differences 
among different cosmological models become larger at higher $z'$s.
Thus, data-points in the HD at high $z'$s
highly discriminate the models, even if the uncertainties are large. (3) Each point in 
the HD translates into a different curve in the space of the cosmological parameters
(degeneration). The wider in $z$ is the data sample, the less elongated along one
curve (less degeneration) will be the parameter confidence levels (Firmani et al. 2007).

\begin{figure*}[!t]
  \includegraphics[width=0.68\columnwidth,height=5.8cm]{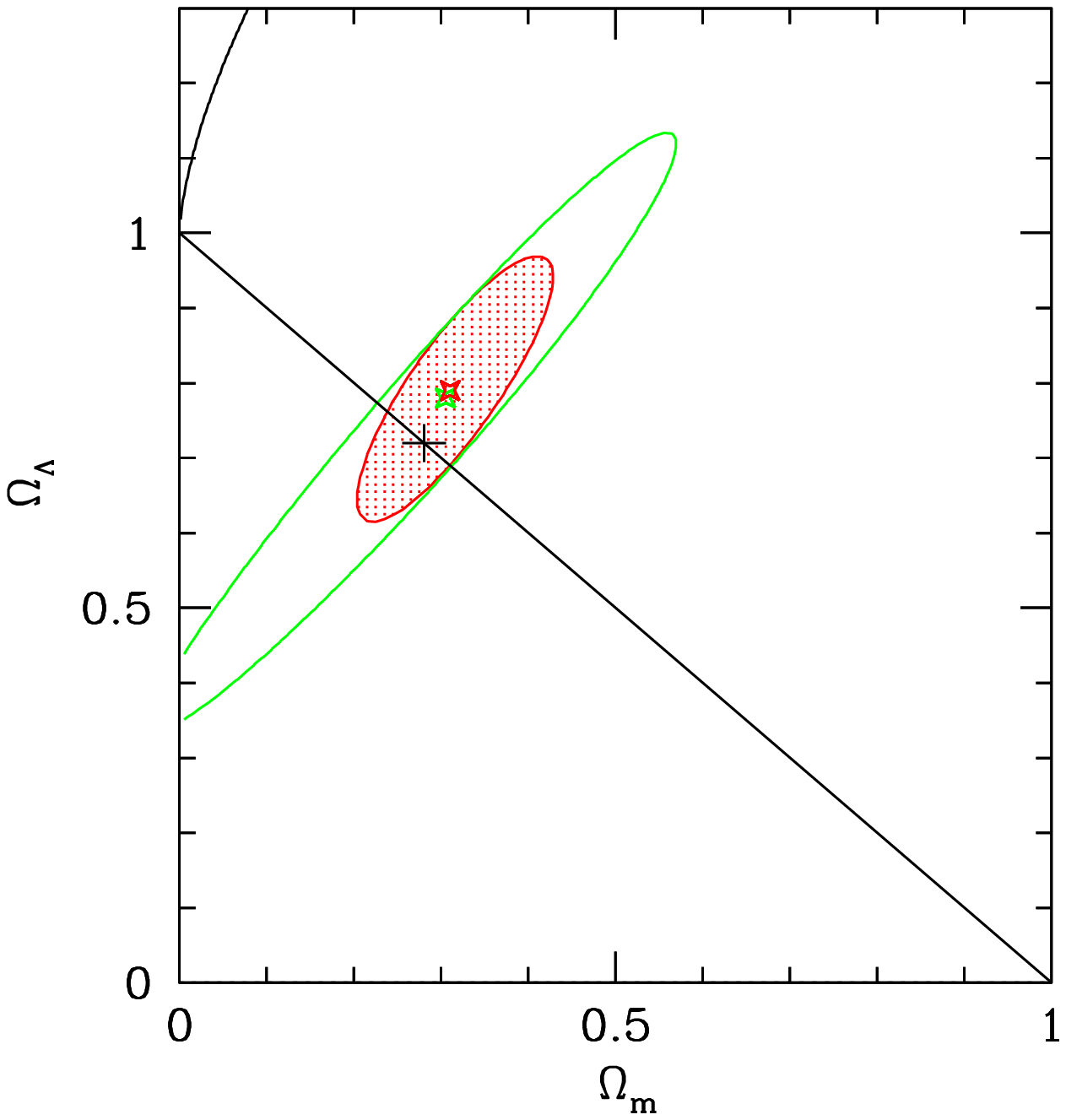}%
  \hspace*{0.4\columnsep}%
  \includegraphics[width=0.68\columnwidth,height=5.8cm]{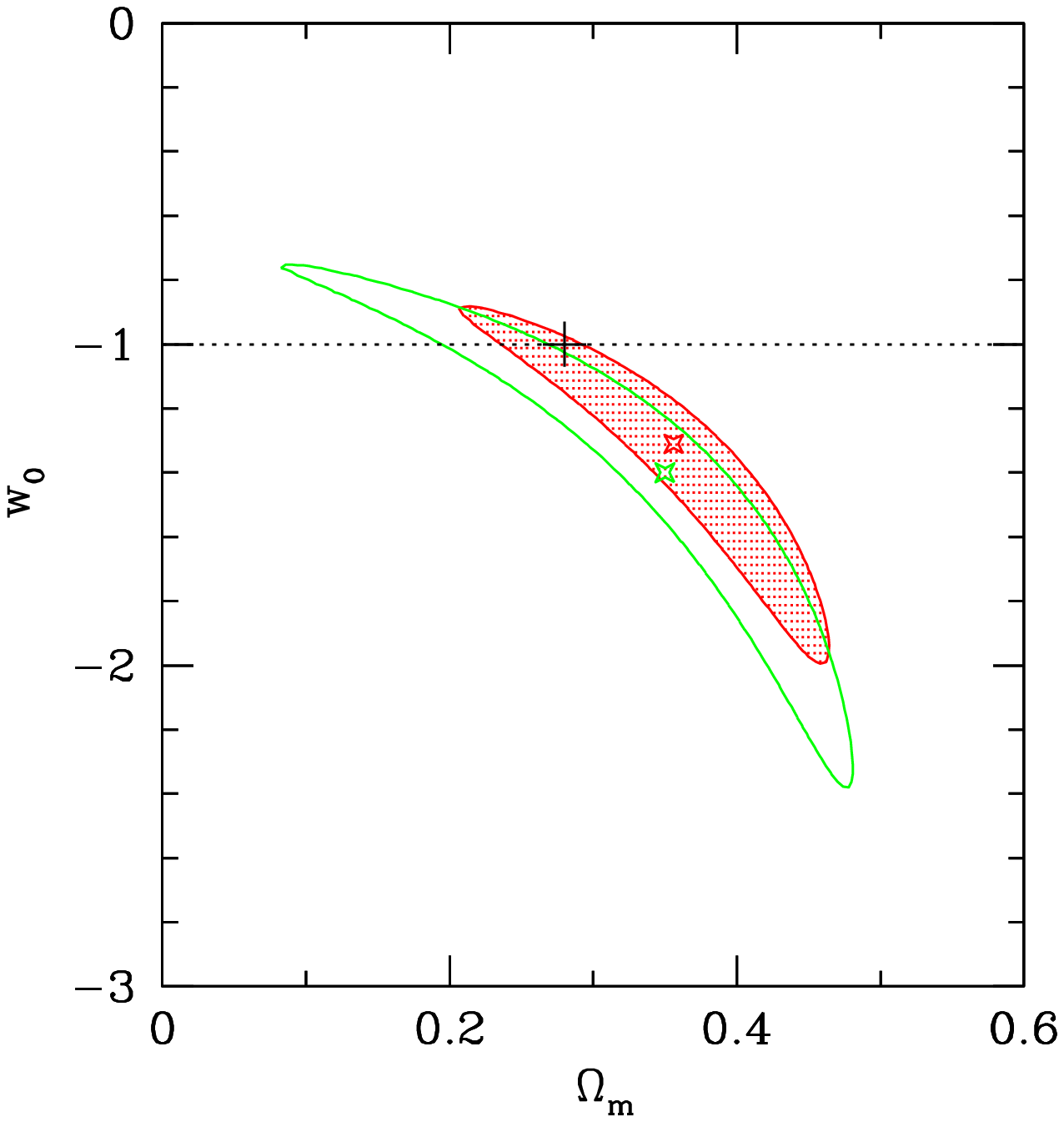}%
  \hspace*{0.4\columnsep}%
  \includegraphics[width=0.68\columnwidth,height=5.8cm]{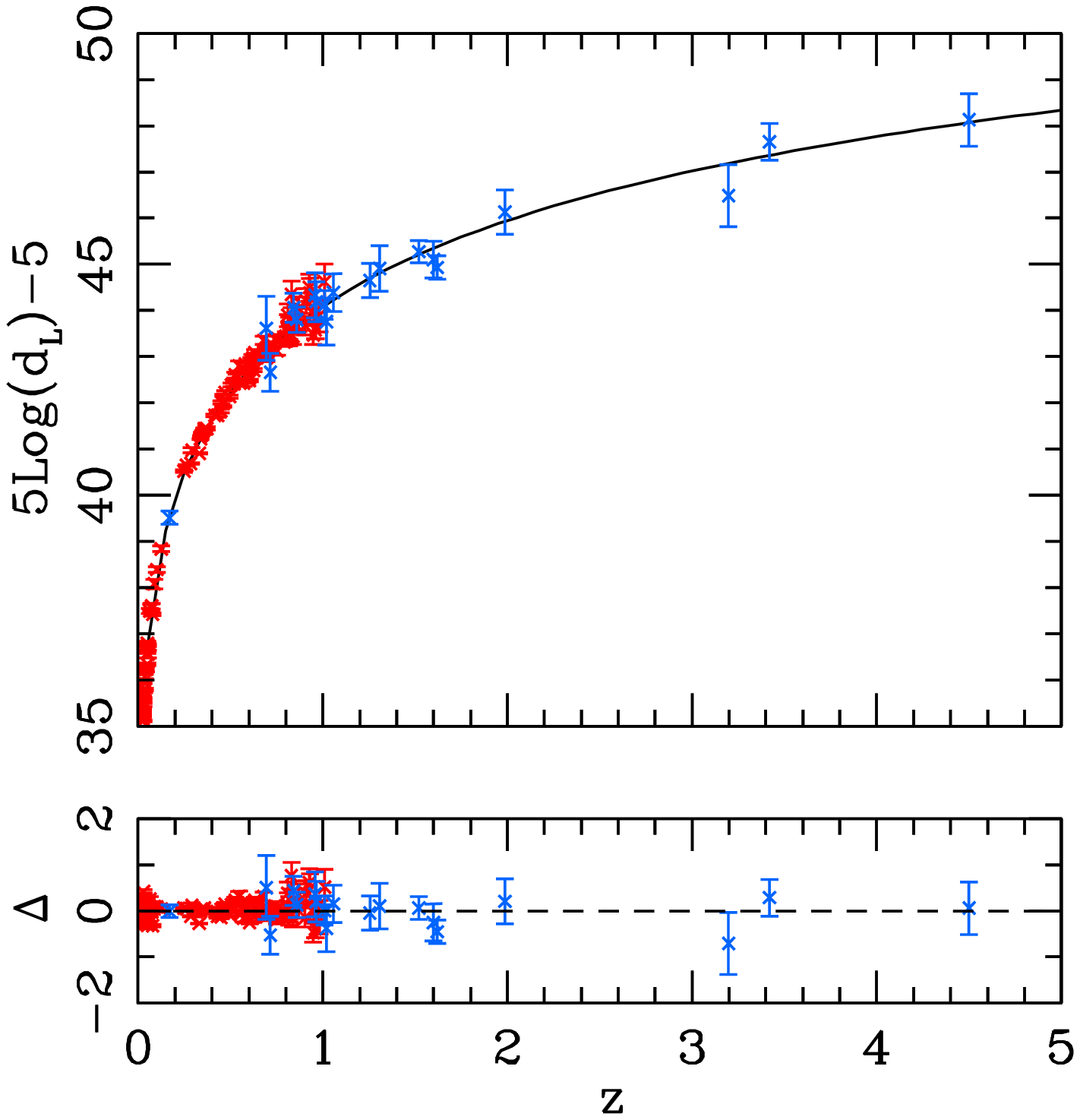}%
  \caption{{\it Left:} Contours at 68.3\% CL's on the (\Om,\OL) plane from the 
SNIa HD using our Bayesian approach to circumvent
the circularity problem (solid green line) and from the combined SNIa+GRB 
HD (red shaded region). {\it Middle:} Same as left panel but
on the (\Om, \wo) plane assuming flatness; \wo~ is the equation of state index,
assumed to be constant. {\it Rigth:} 117 SNIa (red) and 19 GRB (blue)  
data-points in the HD for the flat $\Lambda$CDM model. The residuals of the data-points 
minus the $\Lambda$CDM model are shown in the bottom panel. }
  \label{fig:widefig1}
\end{figure*}

{\bf Caveats.} GRB cosmography is in its infancy and of course there should be
several caveats as was discussed in the literature (e.g., Friedman \& Bloom 2005).
For example, it was argued that when using the ``Ghirlanda'' relation, the results 
are strongly dependent on the assumption about the density distribution of the 
circumbust medium (the dependence of $\theta_j$ on $t_{\rm break}$ changes slightly
with the distribution assumed, and the parameters of one or another distribution
are included in the calculation of $\theta_j$). Despite that the ``Ghirlanda''
correlations are different in one case or another, from the point of view
of cosmography, the results are very similar (Nava et al. 2006). Furthermore,
it was shown that an empirical correlation among \eiso, \ep, and $t_{\rm break}$ holds
(avoiding then the assumption of the circumbust density distribution), which
gives cosmographic results similar to those obtained with the ``Ghirlanda'' correlations
(Liang \& Zhang 2006). Recent \sw~ observations have shown that
the $X-$ray afterglow light curve is more complex than previously though and 
its break time tends to be different from the one inferred in the optical bands.  
Several pieces of evidence suggest that the $X-$ray and optical components come from different
emitting regions; therefore, the requirement that the optical $t_{\rm break}$ should be 
compatible with the $X-$ray one should be relaxed (Nava et al. 2007). 

 Concerning the ``Firmani'' correlation, it was established for prompt $\gamma-$ray 
emission quantities alone. Therefore, it is model independent and does not requires follow--up 
observations. Some of the potential difficulties mentioned in general for the methods of 
standardizing the energetics of GRBs are the systematics and outliers in the correlations, 
the gravitational lensing, the possible evolution of GRB properties, 
and the lack of a physical interpretation of the correlations.  We refer the reader 
to Ghisellini (2007), Firmani et al. (2007), and Ghirlanda (2007) 
for discussions on these caveats. In our opinion, the last problem is the most challenging.
The possibility of evolution is also real (FAGT; Li 2007), but is most likely that this
happens at the level of the overall population and not in what concerns the internal 
emission mechanisms, which control the spectral energy relations.

Finally, it should be said that the current samples of usable GRBs for
the correlations involving energetics are still small. As the samples will increase 
in number, a better treatment of systematics and selection effects will be possible. 
On the other hand, a wide spectral $\gamma-$ray coverage is necessary in order
to obtain reliable correlations involving the energetics of GRBs. Unfortunately,
the \sw~ BAT detector has a too narrow spectral coverage.

In our view, the use of GRBs as cosmological distance indicators has not been sufficiently 
appreciated by the astronomical community. We are aware of the difficulties
of the method, but it should be considered that GRBs offer a {\it unique} 
possibility to constrain the expansion history of the Universe at $z>2$, and in a way that
simply extends (and complements) the method based on SNe type Ia. 
Perhaps we are now in a similar situation as the SN Ia workers in the early 90's, when 
the astronomical community used to react skeptically to their proposals. 
However, the effort is worth it,
because the cosmography with GRBs may offer valuable information to unveil the properties
of what we call Dark Energy, the big mystery of cosmology. 
This mystery stimulates
now the frontiers of physics to move in the direction of exploring new
elements of high energy physics, the unification of gravity and
quantum physics, gravity beyond Einstein relativity, and extra
dimensions.

\section{Outlook}
\label{sec:how-do-tables}

The inference of GFR history and its comparison with the cosmic SFR history will
highly benefit from the growth of the samples of GRBs with accurate $z$ determination.
However, as discussed in \S\S 2.1, the selection effects that plague these samples are 
a challenging issue. Therefore, the indirect methods for inferring the GFR history 
(\S\S 2.3) should be improved in parallel. These methods would largely benefit if large 
samples of GRBs with pseudo--redshifts are constructed. The use of correlations 
among prompt $\gamma-$ray quantities alone (e.g., the ``Firmani'' correlation) is 
the best way for this aim.

The improvement of the known GRB tight correlations or the discovery of new ones,
is not only important for GFR studies but also for cosmography at high $z'$s as
shown in \S 3. The increasing of the observational data is mandatory in this
endeavor. With a sample of $\sim 150$ GRBs, on one hand, the empirical correlations 
might already be calibrated in a small $z$ bin, and in the other one, the HD
would become highly populated as to get tight constraints on the cosmological 
parameters. However, it should be noted that for the correlations that involve
energetics and spectral information, a wide spectral coverage is necessary,
something in which the \sw~ detector fails. The hope is in feature
missions as GLAST.

\acknowledgments

VA thanks the organizing committee for the invitation and for a job very well done.
Support for this work was provided by PAPIIT-UNAM grant IN107706-3.


\begin{thebibliography}

\bibitem[Astier et al.(2006)]{Astier06} Astier, P., et al. 2006, A\&A, 447, 31

\bibitem[Band et al.(1993)]{Band93} Band, D., et al. 1993, ApJ, 413, 281

\bibitem[Bloom(2003)]{b03} Bloom, J.S. 2003, \aj, 125, 2865

\bibitem[Bogomazov et al.(2007)]{2007ARep...51..308B} Bogomazov, A.~I., 
Lipunov, V.~M., \& Tutukov, A.~V.\ 2007, Astronomy Reports, 51, 308 

\bibitem[Bromm \& Loeb(2006)]{2006ApJ...642..382B} Bromm, V., \& Loeb, A.\ 2006, \apj, 642, 382 

\bibitem[Coward et al.(2007)]{2007arXiv0711.0242C} Coward, D.~M., Guetta, 
D., Burman, R.~R., \& Imerito, A.\ 2007,  arXiv:0711.0242 

\bibitem[Dai, Liang \& Xu(2004)]{dai04}
         Dai Z.G., Liang E.W. \& Xu D. 2004, ApJ, 612, L101

\bibitem[Daigne \& Mochkovitch(1998)]{1998MNRAS.296..275D} 
 Daigne, F., \& Mochkovitch, R.\ 1998, \mnras, 296, 275 

\bibitem[Fenimore \& Ramirez-Ruiz(2000)]{2000astro.ph..4176F} Fenimore, E.~E., 
 \& Ramirez-Ruiz, E.\ 2000, arXiv:astro-ph/0004176 

\bibitem[Fiore et al.(2007)]{2007AA...470..515F} Fiore, F., Guetta, D., Piranomonte, S., 
 D'Elia, V., \& Antonelli, L.~A.\ 2007, \aap, 470, 515 

\bibitem[Firmani et al.(2004)]{fagt04} Firmani, C., Avila-Reese, V., Ghisellini, G., \& 
Tutukov, A.~V. 2004, ApJ, 611, 1033 (FAGT)

\bibitem[Firmani et al.(2005)]{fgga05} Firmani, C., Ghisellini, G., Ghirlanda, G., \& Avila-Reese, V. 
2 005, MNRAS, 360, L1

\bibitem[Firmani et al.(2006a)]{Firmani06} Firmani, C., Ghisellini, G., Avila-Reese, V., \& 
Ghirlanda, G. 2006a, MNRAS, 370, 185

\bibitem[Firmani et al.(2006b)]{Firmani06b} Firmani, C., Avila-Reese, V., Ghisellini, G., \& 
Ghirlanda, G., 2006b, MNRAS, 372, L28

\bibitem[Firmani et al.(2007)]{Firmani07} \_\_\_\_\_\_\_\_.  2007, RevMexAA, 43, 203

\bibitem[Friedman \& Bloom(2005)]{2005ApJ...627....1F} 
Friedman, A.~S., \& Bloom, J.~S.\ 2005, \apj, 627, 1 

\bibitem[Fruchter et al.(2006)]{2006Natur.441..463F} 
 Fruchter, A.~S., et al.\ 2006, \nat, 441, 463 
\linebreak\adjustfinalcols

\bibitem[Fynbo et al.(2003)]{2003AA...406L..63F} 
  Fynbo, J.~P.~U., et al.\ 2003, \aap, 406, L63 

\bibitem[Gehrels et al.(2004)]{Gehrels} 
        Gehrels, N. et al. 2004, ApJ, 611, 1005

\bibitem[Ghirlanda(2007)]{2007astro.ph..2212G} Ghirlanda, G.\ 2007, arXiv:astro-ph/0702212 

\bibitem[Ghirlanda et al.(2004a)]{ggl04} Ghirlanda, G., Ghisellini, G., \& Lazzati, D. 2004a, 
 ApJ, 616, 331

\bibitem[Ghirlanda et al.(2004b)]{gglf04} Ghirlanda, G., Ghisellini, G., Lazzati, D., 
\& Firmani, C. 2004b, ApJ, 613, L13





\bibitem[Ghisellini(2007)]{2007MmSAI..78..779G} Ghisellini, G.\ 2007, MmSAI, 78, 779 

\bibitem[Giavalisco et al.(2004)]{g04} Giavalisco, M. et al. 2004, \apjl, 600, L103

\bibitem[Gou et al. 2004(2004)]{gma03} Gou, L. J., M\'esz\'aros, P., Abel, T. \&
Zhang, B. 2004,  ApJ, 604, 508

\bibitem[Guetta 
\& Piran(2007)]{2007JCAP...07....3G} Guetta, D., \& Piran, T.\ 2007, JCAP, 7, 3

\bibitem[Hjorth et al.(2003)]{2003Natur.423..847H} Hjorth, J., et al.\ 
2003, \nat, 423, 847 

\bibitem[Hopkins 
\& Beacom(2006)]{2006ApJ...651..142H} Hopkins, A.~M., \& Beacom, J.~F.\ 2006, \apj, 651, 142 

\bibitem[Kistler et al.(2008)]{2008ApJ...673L.119K} Kistler, M.~D., 
Y{\"u}ksel, H., Beacom, J.~F., \& Stanek, K.~Z.\ 2008, \apjl, 673, L119 

\bibitem[Le 
\& Dermer(2007)]{2007ApJ...661..394L} Le, T., \& Dermer, C.~D.\ 2007, \apj, 661, 394 

\bibitem[Le Floc'h et al.(2006)]{2006ApJ...642..636L} Le Floc'h, E. et al. 2006, \apj, 642, 636 

\bibitem[Li(2007)]{2007MNRAS.379L..55L} Li, L.-X.\ 2007, \mnras, 379, L55 

\bibitem[Li et al.(2006)]{2006astro.ph.12060L} Li, H., Su, M., Fan, Z., 
Dai, Z., \& Zhang, X.\ 2006, arXiv:astro-ph/0612060 

\bibitem[Liang \& Zhang(2006)]{liang05} Liang,  E., \& Zhang, B. 2006, ApJ, 633, 611

\bibitem[Lloyd-Ronning, Fryer \& Ramirez-Ruiz(2002)]{Lloyd02} 
     Lloyd-Ronning, N.~M., Fryer, C.~L., \& Ramirez-Ruiz, E.\ 2002, ApJ, 574, 554 

\bibitem[Meszaros(2006)]{2006RPPh...69.2259M} Meszaros, P.\ 2006, RPPh, 69, 2259 

\bibitem[Nava et al.(2006)]{Nava06} Nava, L., Ghisellini, G., Ghirlanda, G., Tavecchio, F., 
\& Firmani,  C. 2006, A\&A, 450, 471

\bibitem[Nava et al.(2007)]{2007MNRAS.377.1464N} Nava, L., Ghisellini, G., 
Ghirlanda, G., Cabrera, J.~I., Firmani, C., 
\& Avila-Reese, V.\ 2007, \mnras, 377, 1464 

\bibitem[Nuza et al.(2007)]{2007MNRAS.375..665N} Nuza, S.~E. et al. 2007, \mnras, 375, 665  

\bibitem[Piran(2005)]{2005RvMP...76.1143P} Piran, T.\ 2005, Reviews of 
Modern Physics, 76, 1143 

\bibitem[Porciani \& Madau(2001)]{Porciani00} Porciani, C., \& Madau, P. 2001, ApJ, 548, 522 

\bibitem[Priddey et al.(2006)]{2006MNRAS.369.1189P} Priddey, R.~S. et al. \ 2006, \mnras, 369, 1189  

\bibitem[Savaglio(2006)]{2006NJPh....8..195S} Savaglio, S.\ 2006, New 
Journal of Physics, 8, 195 

\bibitem[Schaefer(2007)]{schaefer} Schaefer, B. E. 2007, ApJ, 660, 16

\bibitem[Stanek et al.(2003)]{2003ApJ...591L..17S} Stanek, K.~Z., et al.\ 
2003, \apjl, 591, L17

\bibitem[Stanek et al.(2006)]{2006AcA....56..333S} Stanek, K.~Z., et al.\ 
2006, Acta Astronomica, 56, 333 

\bibitem[Xu, Dai, \& Liang(2005)]{XDL05} Xu, D., Dai, Z.~G., \& Liang, E.~W. 2005, ApJ, 633, 603

\bibitem[Yoon et 
al.(2006)]{2006AA...460..199Y} Yoon, S., Langer, N., \& Norman, C.\ 2006, \aap, 460, 199

\bibitem[Zhang(2007)]{2007ChJAA...7....1Z} Zhang, B.\ 2007, ChJAA, 7, 1 

\bibitem[Zhang 
\& M{\'e}sz{\'a}ros(2004)]{2004IJMPA..19.2385Z} Zhang, B., \& M{\'e}sz{\'a}ros, P.\ 2004, 
IJMPA, 19, 2385 

\end{thebibliography}
\end{document}